\documentclass[twoside]{article}

\usepackage{lipsum} 

\usepackage[sc]{mathpazo} 
\usepackage[T1]{fontenc} 
\usepackage{microtype} 

\usepackage[hmarginratio=1:1,top=32mm,columnsep=20pt]{geometry} 
\usepackage{multicol} 
\usepackage[hang, small,labelfont=bf,up,textfont=it,up]{caption} 
\usepackage{booktabs} 
\usepackage{float} 
\usepackage{hyperref} 
\usepackage{natbib} 
\usepackage{graphicx}
\usepackage{caption} 
\usepackage{subcaption} 
\usepackage{lettrine} 
\usepackage{paralist} 

\usepackage{abstract} 

\usepackage{titlesec} 
\renewcommand\thesection{\Roman{section}} 
\renewcommand\thesubsection{\Roman{subsection}} 
\titleformat{\section}[block]{\large\scshape\centering}{\thesection.}{1em}{} 
\titleformat{\subsection}[block]{\large}{\thesubsection.}{1em}{} 

\usepackage{fancyhdr} 
\title{\vspace{-15mm}\fontsize{24pt}{10pt}\selectfont\textbf{Beyond the New Horizon: The Future of Pluto}} 

\author{
\large
\textsc{Michael B. Lund}\\
\normalsize Vanderbilt University \\ 
\normalsize \href{mailto:michael.b.lund@vanderbilt.edu}{michael.b.lund@vanderbilt.edu} 
\vspace{-5mm}
}
\date{}

\begin{document}

\maketitle 

\thispagestyle{fancy} 


\begin{abstract}

\noindent Since its discovery in 1930, Pluto's mass has been a value that has repeatedly been calculated. Additionally, the search for Planet X prior to Pluto's discovery results in mass calculations that date back several decades earlier. Over its observed history, the mass of Pluto has consistently decreased. We reassess earlier predictions of Pluto's fate, and rule out the hypothesis that Pluto's mass has been constant over the last century. We are able to fit linear and quadratic equations to Pluto's mass as a function of both time and distance. The observations that will be made by New Horizons will help to determine if we can expect Pluto to continue to shrink until it has negative mass, or if it will begin to increase in mass again.

\end{abstract}


\begin{multicols}{2} 

\section{Introduction}
\lettrine[nindent=0em,lines=3]{T} he search for planets in the outer solar system has been an active endeavor in astronomy, especially after the discovery of the first planet to be added to the classical planets with the discovery of Uranus in 1781 by William Hershel. The discovery of Neptune in 1851 by  Johann Gottfried Galle and Heinrich Louis d'Arrest in 1846, using predictions made by Urbain Le Verrier, demonstrated the ability to discover objects in the solar system indirectly by locating their effect on already known planets. The most controversial object, both in the scientific and public realm, would be the discovery of Pluto in 1930. Prior to this time, the existance of an additional planet had been suggested by multiple authors and the search for Planet X was well under way, including predictions of its mass \citep{Pickering1909, Lowell1915}. This provides an over 80 year baseline of observations post-discovery, and even more if the inferred masses are included in this analysis.

Over this time, the mass calculated for Pluto has changed from estimates on the order of the mass of the earth around the time of discovery to the current mass estimates that are only a fraction of the earth's mass. Here, we consider that this data may still be useful and worthy of consideration. The idea that Pluto could be shrinking is not new \citep{Disney1935, Dessler1980}, and we revisit that concept to see how more recent observations help us constrain Pluto's evolution over the last century and what this can tell us about the origins of Pluto.

\section{Data}
With a long history of mass calculations, we look at 75 years of published masses for Pluto (1931 to 2006), as well as including two mass calculations for Planet X. We supplement this data with heliocentric distance calculations for Pluto that have been gathered with the Stellarium software\footnote{Available from \url{http://www.stellarium.org}}. Our full data set is in Table~\ref{table:PlutoData}.
\footnotetext[2]{Sourced from \citet{Weintraub2014}}
\begin{table*}[!htb]
\caption{Pluto Masses and Distances}
\label{table:PlutoData}
\centering
\begin{tabular}{llll}
\toprule
Year & Distance (AU) & Mass ($M_{earth}$)    & Citation   \\
\midrule
1909 & 45.6          & 2            & \citet{Pickering1909}  \\
1915 & 44.52         & 6.67         & \citet{Lowell1915}     \\
1931 & 41.09         & 0.94         & \citet{Nicholson1931}  \\
1931 & 41.09         & 0.75         & \citet{Pickering1931}  \\
1942 & 38.38         & 0.91         & Wylie\footnotemark[2]    \\
1949 & 36.58         & 0.8          & Kuiper\footnotemark[2]   \\
1968 & 32.06         & 0.18         & \citet{Duncombe1968}   \\
1971 & 31.49         & 0.11         & \citet{Seidelmann1971} \\
1989 & 29.66         & 0.0021655009 & \citet{Binzel1989}     \\
1993 & 29.72         & 0.00219357   & \citet{Null1993}       \\
1997 & 29.95         & 0.0022069658 & \citet{Foust1997}      \\
2006 & 31.05         & 0.0021851488 & \citet{Buie2006}      \\

\bottomrule
\end{tabular}
{\\Masses and heliocentric distances for Pluto over the last century.}
\end{table*}
\section{Analysis}
We consider two simple possible functions in this work; the first is that the mass is a function of time, and the second is that the mass is a function of heliocentric distance. The conventional model we include is that the mass of Pluto over the last century has actually been a constant. However in addition to this, we consider that this relation may be linear, quadratic, or a power law. We also include the proposed solution from \citet{Dessler1980}, which is provided in Equation~\ref{eq:DR}:
\begin{equation}\label{eq:DR}
  M = 12 [cos\frac{\pi (t-1848)}{272}]^{\pi}
\end{equation}

We also look at the validity of using the Planet X values for this fit, and so we attempt to fit the data using both the data from 1909 and 1915, and only the data following the official discovery of Pluto.

\subsection{Mass as a Function of Time}
\begin{figure*}[!htb]
  \begin{center}
    \begin{subfigure}[b]{0.45\textwidth}
      \includegraphics[width=\textwidth]{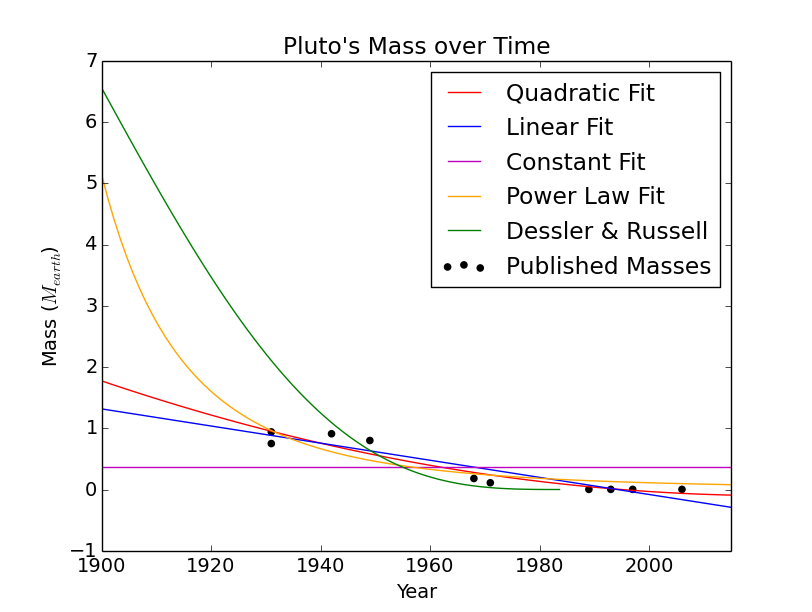}
    \end{subfigure}
    \begin{subfigure}[b]{0.45\textwidth}
      \includegraphics[width=\textwidth]{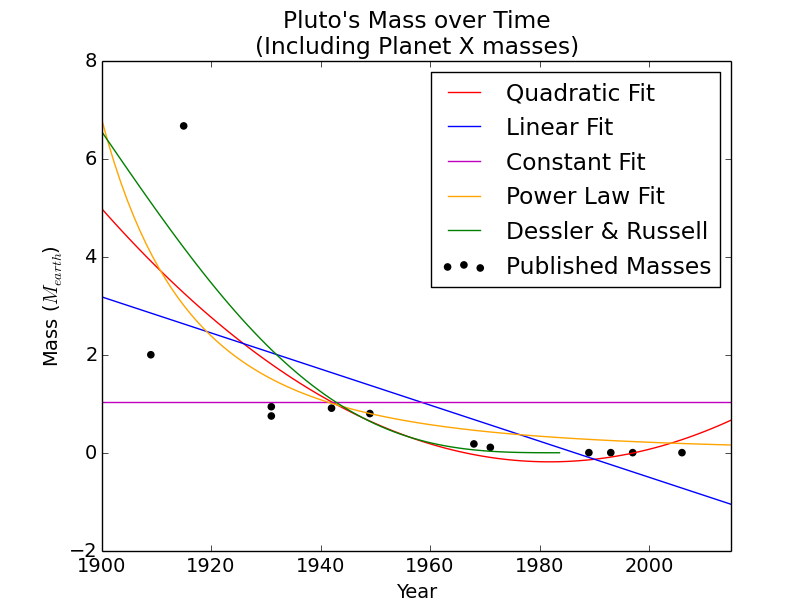}
    \end{subfigure}
  \caption{Best fit lines for functions as determined by the Pluto masses on left, and Pluto and Planet X masses on the right.}
  \label{fig:MassTimeT}
  \end{center}
\end{figure*}

We conduct an analysis of the mass as a function of time by attempting to fit to the data both with and without the Planet X mass values. We display these two fits in Figure~\ref{fig:MassTimeT}. With each of these fits, we then also calculate the values for the $\chi^{2}$ and \emph{p} values for each of these models, displayed in Table~\ref{table:MassTime1} and Table~\ref{table:MassTime2}.

\begin{table*}[!htb]
\caption{Quality of Fits to Pluto-only Data}
\label{table:MassTime1}
\centering
\begin{tabular}{lllll}
\toprule
Model  & Constant Value & Linear Fit          & Quadratic Fit & Power Law Fit \\
\midrule
$\chi^{2}$   & 4.3004         & 0.2153              & 0.2461              & 0.9861        \\
\emph{p}-value & 0.8290         & \textgreater 0.9999	& 0.9997        & 0.9868       \\
\bottomrule
\end{tabular}
\end{table*}

\begin{table*}[!htb]
\caption{Quality of Fits to Pluto \& Planet X Data}
\label{table:MassTime2}
\centering
\begin{tabular}{lllll}
\toprule
Model  & Constant Value & Linear Fit          & Quadratic Fit & Power Law Fit \\
\midrule
$\chi^{2}$   & 37.5450         & 7.7022              &      3.9001         & 7.4395        \\
\emph{p}-value & \textless 0.0001         & 0.5644	& 0.8660        & 0.4900        \\
\bottomrule
\end{tabular}
\end{table*}

We don't compare the function that was proposed by \citet{Dessler1980}, as their proposal was that Pluto would have an imaginary mass after 1984, which appears to be inconsistent with the real masses that have been measured in the last 20 years. However, the $\chi^{2}$ is not a value we can calculate, but we feel that further indicates that this model can be discarded.

In both cases, we find that the least likely model for the observed mass history of Pluto is that Pluto has had a constant mass. We find that linear, quadratic, and power-law fits are all robust in the Pluto-only data, however when we include the Planet X data as well, the quadratic fit may be the most likely. It is particularly persuasive that the quadratic fit we find when we include the Planet X masses appears to have a minimum around the same point where \citet{Dessler1980} had also predicted that Pluto's mass would drop below zero. The starkest difference will be whether or not the small increase we see in Pluto's mass from the early 1990s \citep{Binzel1989, Null1993} to the late 1990s and early 2000s \citep{Foust1997, Buie2006} continues into the future as a quadratic function would predict, or if this is an aberration in an otherwise linear trend.

\subsection{Mass as a Function of Distance}
\begin{figure*}[!htb]
  \begin{center}
    \begin{subfigure}[b]{0.45\textwidth}
      \includegraphics[width=\textwidth]{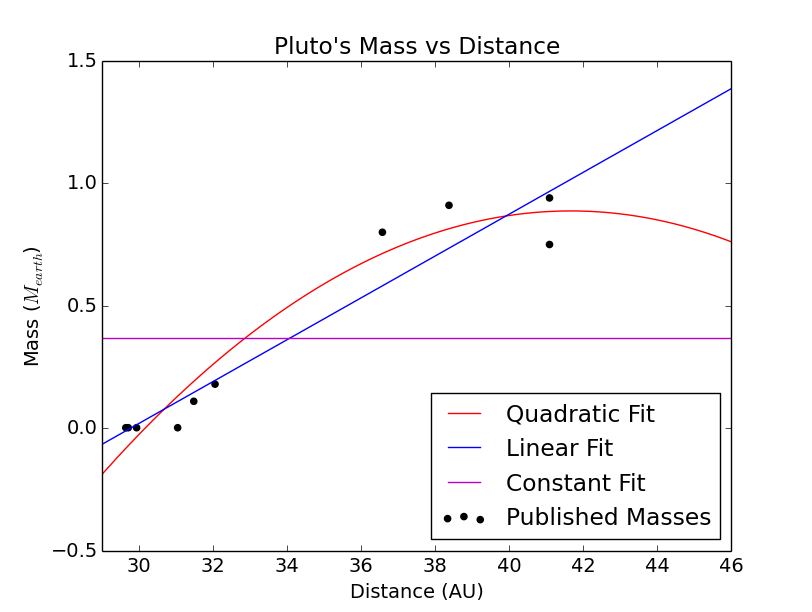}
    \end{subfigure}
    \begin{subfigure}[b]{0.45\textwidth}
      \includegraphics[width=\textwidth]{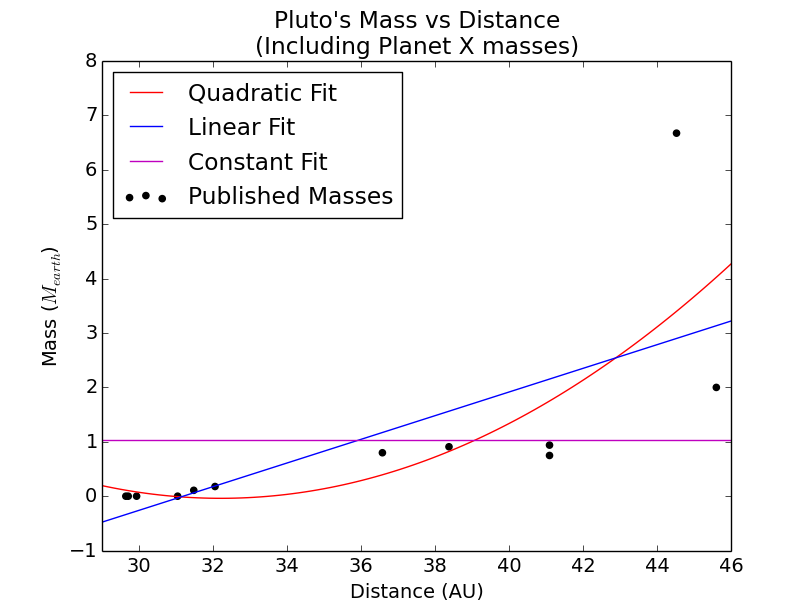}
    \end{subfigure}
  \caption{Best fit lines for functions as determined by the Pluto masses on left, and Pluto and Planet X masses on the right.}
  \label{fig:MassDistanceT}
  \end{center}
\end{figure*}

For our analysis of mass as a function of distance, we are not able to get suitable fits for the Power Law fit, and so we omit that equation. Additionally, the Dessler-Russell equation was specifically set to be a function of time, and so as it isn't applicable here we exclude it as well, limiting our consideration to only the constant value, linear, and quadratic fits. We display these fits in Figure~\ref{fig:MassDistanceT}. With each of these fits, we then also calculate the values for the $\chi^{2}$ and \emph{p} values for each of these models, displayed in Table~\ref{table:MassDistance1} and Table~\ref{table:MassDistance2}.

\begin{table*}[!htb]
\caption{Quality of Fits to Pluto-only Data}
\label{table:MassDistance1}
\centering
\begin{tabular}{lllll}
\toprule
Model  & Constant Value & Linear Fit          & Quadratic Fit\\
\midrule
$\chi^{2}$   & 4.3004         & 0.2725              & 0.0337 \\
\emph{p}-value & 0.8290         & \textgreater 0.9999 & \textgreater 0.9999 \\
\bottomrule
\end{tabular}
\end{table*}

\begin{table*}[!htb]
\caption{Quality of Fits to Pluto \& Planet X Data}
\label{table:MassDistance2}
\centering
\begin{tabular}{lllll}
\toprule
Model  & Constant Value & Linear Fit          & Quadratic Fit\\
\midrule
$\chi^{2}$   & 37.5450         & 6.3603              &      3.7987\\
\emph{p}-value & \textless 0.0001         & 0.7034 & 0.8748\\
\bottomrule
\end{tabular}
\end{table*}

Again, we see that we get better fits for the linear and quadratic equations than we get for Pluto as a constant value. Beyond that, we do see that there is a very good relation between the mass and the distance, in this case stronger (although not statistically more significant) than what we observed for the mass as a function of time. It is worth noting that a distance-dependent mass does provide a natural limit on the mass, preventing it from becoming smaller than Pluto was at perihelion (and potentially prohibiting negative masses) as well as an upper limit when Pluto is at aphelion (removing the possibility of a mass runaway in the future).

\section{Discussion}
The strong evidence we provide as an indication that Pluto has changed mass over the last century gives a very natural explanation as to why the status of Pluto has been of such great debate over the last decade. It remains to be seen if we are observing Pluto at a local minima, in which case it may increase to planet status again in the future, or if it will continue to shrink until such a point as its planet status is irrelevant or otherwise unquestioned. The possibility that Pluto's mass may increase in the future is not a result that was considered by \citet{Dessler1980}, however the predictions that could be made from a linear fit for Pluto's mass could give somewhat similar predictions of the disappearance of Pluto (or the realization of a negative mass) somewhere in the near future.

The physical meaning of a distance-dependant mass would need further analysis to understand, and so we leave that to further contributions from the community. This solution is of note, however, as it will provide a minimum and maximum mass for Pluto that eliminates negative or infinite masses. We can better address the question of time-dependent mass-loss, as this is not an unknown phenomenon. It has been seen in high mass planets already, such as the large mass loss that has been observed in WASP-43b\citep{Czesla2013}. There are also measurements made for the mass-loss of solar system bodies, including Pluto itself \citep{Johnson2015} and comets through multiple physical channels \citep{Napier2015}.

We can further surmise that as the mass has decreased over time, we can also imagine that prior to the speculation of Planet X, that Pluto's mass was already decreasing. As it would be unrealistic to presume that Pluto had a very large mass in the mass (on the order of Neptune) without causing Neptune's orbit to exhibit something distinctly non-circular, this would imply that Pluto has not been in its current orbit until relatively recently. This would be consistent with the idea that Pluto's orbit is still not stable, and so is undergoing some scale of evolution \citep{SUSSMAN1988}. Further constraining this function will allow us to infer more about the dynamic history of Pluto.

\section{Summary}
While we are in broad agreement with \citet{Dessler1980} that Pluto has decreased in mass over the last century, we don't find their proposal for the function that best represents this mass loss. Our best indications are that the mass loss should be represented by a linear or quadratic fit. New Horizons will help provide great insight to this, as a linear fit would indicate that Pluto would have reached zero mass somewhere after 1994, and would be continuing to shrink. The quadratic fit for the near future, when we exclude Planet X, would similarly be negative for the near future. The New Horizons observations that will be made by Pluto in the next year will be very useful in constraining which of these functional forms is most accurate, and may provide further evidence that Pluto's mass is distance-dependent rather than time-dependent.


\bibliographystyle{apalike}
\bibliography{article_2}


\end{multicols}

\end{document}